\documentstyle[a4,aps,prb,graphics]{revtex}

\begin{document}

\title{ 
Density-matrix renormalization study of the Hubbard model on a Bethe lattice
      }

\author{Marie-Bernadette Lepetit, Maixent Cousy and G.\ M.\ Pastor}

\address
{Laboratoire de Physique Quantique, Unit\'e Mixte de Recherche 5626 du CNRS,\\
118 route de  Narbonne, F-31062 Toulouse, France}

\date{\today}
\maketitle

\begin{abstract} 

The half-filled Hubbard model on the Bethe lattice with coordination number
$z=3$ is studied using the density-matrix renormalization group (DMRG) 
method. Ground-state properties such as the energy $E$, average 
local magnetization $\langle \hat S_z \rangle$, its fluctuations
$\langle \hat S_z^2 \rangle - \langle \hat S_z\rangle^2$ and 
various spin correlation functions 
$\langle \hat S_z(i) \hat S_z(j) \rangle - 
\langle S_z(i)\rangle \langle S_z(j) \rangle$
are determined as a function of the Coulomb interaction strength $U/t$. 
The calculated local magnetic moments $\langle \hat S_z(i)\rangle$ 
increase monotonically with increasing Coulomb repulsion $U/t$ 
forming an antiferromagnetic spin-density-wave state which matches
the two sublattices of the bipartite Bethe lattice.
At large $U/t$, $\langle \hat S_z(i)\rangle$ is strongly reduced 
with respect to the saturation value $1/2$ due to
exchange fluctuations between nearest neighbors (NN) spins 
($|\langle S_z(i)\rangle|\simeq 0.35$ for $U/t\to +\infty$).
$\langle S_z(i)^2\rangle - \langle S_z(i)\rangle^2$ 
shows a maximum for $U/t=2.4$--$2.9$ which results from the 
interplay between the usual increase of $\langle S_z(i)^2\rangle$
with increasing $U/t$ and the formation of important permanent moments 
$\langle S_z(i)\rangle$ at large $U/t$. NN sites show 
antiferromagnetic spin correlations which increase with increasing 
Coulomb repulsion. In contrast next NN sites are very weakly correlated 
over the whole range of $U/t$. 
The accuracy of the DMRG results is discussed by comparison 
with tight-binding exact results, independent DMRG calculations for 
the Heisenberg model and simple first-order perturbation estimates.

\end{abstract} 

\vspace{0.5cm}

\section{Introduction}
\label{sec:Intro}

Bethe lattices or Cayley trees have often been the basis of
very attractive models for the theoretical study of various properties 
of solids. A Bethe lattice is completely characterized by the number 
of nearest neighbors $z$ and by the lack of closed loops. 
The later feature simplifies calculations considerably allowing 
in many cases to obtain useful insights on the physics of complex problems, 
for example, in theory of many electron systems or in the theory of alloys 
and other disordered systems. Recently, the interest in the study of 
strongly correlated fermions on Bethe lattices has been renewed by the 
advances achieved in the limit of infinite spatial dimensionality 
($d\to\infty$) \cite{voll}. Indeed, this  lattice  provides a
systematic mean of realizing the $d=\infty$ limit by letting $z\to \infty$ 
and scaling the nearest neighbor (NN) hopping integrals 
as $t=W/(4\sqrt{z})$, where $W$ refers to the band width. In this context,
an important research effort has been dedicated to the half-filled 
Hubbard model. This concerns mainly the metal-insulator transition 
within the paramagnetic phase and also the properties of the 
antiferromagnetic phase which in the absence of frustrations
is the most stable solution at low temperatures  (bipartite lattice) 
\cite{kotliar,georges,hong}.
Therefore, it would be of considerable interest to determine the 
properties of the half-filled Hubbard model on a Bethe lattice with  
finite $z$ by using accurate numerical methods. 
Such numerical studies could be very useful, particularly in view 
of the possibility of improving the $z=\infty$ 
equations by introducing $1/z$ corrections.

The aim of this paper is to determine several ground-state properties
of the half-filled Hubbard model on a $z=3$ Bethe lattice as a function 
of the Coulomb interaction strength $U/t$. For this purpose we take 
advantage of a property that Bethe lattices share with one-dimensional (1D)
chains, namely, the fact that there is only one path 
between any pair of sites in the system. This characteristic allows 
the design of a simple real-space renormalization scheme 
in order to apply the density-matrix renormalization group 
(DMRG) method \cite{dmrg0}. Density-matrix renormalization is a very 
powerful technique which was first proposed a few years ago in the
context of 1D spin systems. Since then it has 
been rapidly extended to become one of the leading numerical tools 
for the study of low-dimensional correlated quantum systems
including recently two-dimensional lattices of finite size \cite{dmrg1}. 
The success and wide range of applications found by this approach 
rely on two main qualities: its high accuracy even for systems as large 
as a few hundreds of sites, which allows safe extrapolations to 
the thermodynamic limit, and its flexibility concerning the 
model Hamiltonian under study (Heisenberg, $t-J$,
Hubbard, Kondo, etc). However, except for the work of 
Otsuka on the spin-1/2 $XXZ$ Hamiltonian \cite{otsuka}, 
the infinite DMRG calculations have always been limited to 1D problems. 
To our knowledge, this is the first time that the DMRG algorithm for 
infinite systems is applied to a fermion model not having the 1D topology.

The remainder of the paper is organized as follows. In the next section 
the main details of the application of the DMRG method to the Bethe 
lattice are given. 
Results for the ground-state energy and several spin and charge local 
properties are presented and discussed in Sec.~\ref{sec:res}. 
Finally, Sec.~\ref{sec:conc} summarizes our conclusions.

\section{Details of the calculation}
\label{sec:teo}

\begin{figure}[x]
\vspace{1cm}
\centerline{\resizebox{4cm}{4cm}{\includegraphics{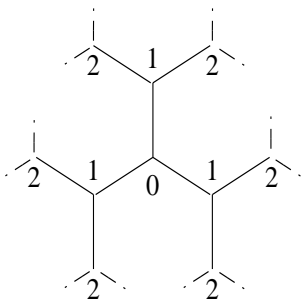}}}
\vspace{5mm}
\caption{Illustration of the Bethe lattice with coordination number $z=3$. 
The numbers label non-equivalent sites.}
\label{fig:bethe-lat1}
\end{figure}

We consider the Hubbard Hamiltonian \cite{hub}
\begin{equation}
H = -t\; \sum_{\langle i,j\rangle, \sigma}\; 
         \hat c^\dagger_{i\sigma} \; \hat c_{j\sigma} \; 
+ \;U\; \sum_i \;\hat n_{i\uparrow} \;\hat  n_{i\downarrow} 
\label{ham}
\end{equation}
on a Bethe lattice with coordination number $z=3$ 
(see Fig.~\ref{fig:bethe-lat1}). In the usual notation, 
$\hat c^\dagger_{i\sigma}$ ($\hat c_{i\sigma}$) creates (annihilates) 
an electron with spin $\sigma$ at site $i$,
$\hat n_{i\sigma} = \hat c^\dagger_{i\sigma} \hat c_{i\sigma}$ is the
number operator, $t$ refers to the nearest neighbor (NN) 
hopping integral, and $U$ to the on-site Coulomb repulsion.
Several ground-state properties of the Hubbard model are determined
using the DMRG method \cite{dmrg0}. This is an iterative projection 
technique which allows to include the most relevant part of the 
ground-state wave function on a limited number of many-body states.
The system is partitioned into several regions in real space or
blocks between which renormalized interactions are computed. 
The accuracy of the method is controlled 
by the number of states $m$ retained for the description of each block. 
In the DMRG algorithm for infinite systems, the size of the system 
is increased at each renormalization 
step and the properties of the thermodynamic limit are determined 
by extrapolating the succession of finite system calculations.

\begin{figure}[x]
\vspace{1cm}
\centerline{\resizebox{12cm}{4.5cm}{\includegraphics{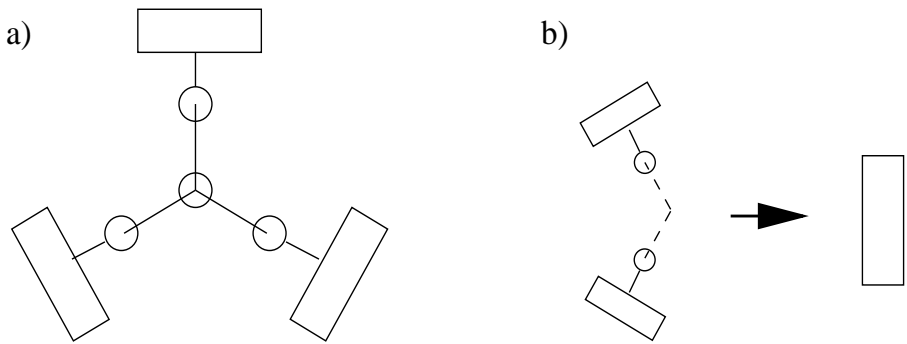}}}
\vspace{5mm}
\caption{
(a) Renormalized Bethe lattice ($z=3$, see Fig.~\protect\ref{fig:bethe-lat1}) 
and (b) superblock renormalization procedure ($\nu \ge 3$). Circles 
represent sites which are treated exactly and rectangles renormalized 
blocks. Notice that the total number of sites $N_a=3\times 2^{\nu} - 2$ 
increases exponentially with the number of renormalization iterations $\nu$.
        }
\label{fig:rg}
\end{figure}

The renormalization procedure used for the $z=3$ Bethe lattice is 
illustrated in Fig.~\ref{fig:rg}. Starting from a central site $i=0$
and the two shells of its first and second NN's, a new 
shell of NN's is added at each iteration $\nu$. The total number sites 
is thus given by $N_a = 3\times 2^\nu - 2$. Notice the contrast with 
the usual DMRG scheme for 1D systems where $N_a$ increases
linearly with $\nu$ \cite{dmrg0}. 
The central site and its NN's (the sites on the first shell) are treated
exactly at all iterations and the renormalizations are performed on 
the two branches which start at each of the first-shell sites 
(see Fig.~\ref{fig:rg}). Notice that no renormalization is actually done 
until $\nu=3$, i.e., the 10-site system with two sites in each 
block is treated exactly. In practice, the extrapolations of the 
considered properties to $\nu\to\infty$ converge after 
$\nu\simeq 15$ iterations. At this point $N_a\simeq 2\times 10^{5}$.
It should be however noted that in a Bethe lattice the surface to volume 
ratio does not vanish with increasing number of shells $L$. 
Taking into account that the number of sites in the $l$th shell is 
$N_s = z (z-1)^{l-1}$ ($l\ge 1$), one finds for $z=3$ that 
1/2 of the sites belong to the outermost shell, 1/4 to the first shell 
below the surface, 1/8 to the second shell below the surface and so forth.
Consequently, global properties of the system such the average 
ground-state energy per site $E_s$ are dominated by the outermost shells. 
Bulk properties corresponding to the translational invariant situation
have to be calculated locally. For instance, the ground-state energy 
per site $E$ is determined from 
$E= U \langle  \hat n_{0\uparrow} \hat n_{0\downarrow}\rangle +
(zt/2) \sum_\sigma \langle \hat c_{1\sigma}^\dagger \hat c_{0\sigma} +
\hat c_{0\sigma}^\dagger \hat c_{1\sigma}\rangle$
by using the density matrix reduced to the central sites $i=0$ and $i=1$
(see Fig.~\ref{fig:bethe-lat1}).
In Sec.~\ref{sec:res} it will be shown that the 
present DMRG algorithm yields accurate results for both global properties
including the surface and local bulk-like properties.

For the calculations we take advantage of a theorem by Lieb which states 
that for $U>0$ the ground-state spin $S$ of the half-filled Hubbard 
Hamiltonian on a bipartite lattice is $S=|N_A - N_B|/2$,
where $N_A$ and $N_B$ are the number of sites belonging to the two 
sublattices $A$ and $B$ ($N_a = N_A + N_B$ even) \cite{liebS}.
In the Bethe lattice considered in this paper,
even and odd shells constitute the two sublattices $A$ and 
$B$ (NN hopping). At the renormalization iteration $\nu$ the
ground-state spin is $S = 2^{\nu-1}$ so that the calculations can be 
performed in the subspace of maximal $S_z = S$. This exact result 
provides an important simplification which appears to be crucial in 
order to achieve reliable results with present computer 
facilities, since for $S_z = S$ the Hilbert space is much smaller than 
for $S_z = 0$. Thus, the number of states $m$ kept in the
renormalized blocks can be reduced significantly without loss of accuracy.
For example for $U=0$, we have performed DMRG calculations by using $S_z=0$ 
and $S_z=S$, keeping in both cases the same number of states $m=20$ in the 
renormalized blocks. After the first few iterations one finds important 
differences in the ground-state energy per site, the $S_z=0$ results being
$2.6\times 10^{-2}t$ higher than those corresponding to $S_z=S$. 
Moreover, the sum $P_m$ of the retained eigenvalues of the density matrix 
--- a good criterion to estimate the quality of a DMRG 
calculation \cite{dmrg0} --- follows the same trend. Indeed, for $S_z=S$, 
$1-P_m$ is always smaller than $7\times 10^{-4}$ while for $S_z=0$,
$1-P_m$ can be as large as $2\times 10^{-2}$, a value which in practice is 
too large for obtaining accurate results. In addition it should be noted
that the uncorrelated limit is the most difficult case in DMRG 
calculations. $1-P_m$ is in fact always 
smaller for finite $U/t$ than for $U=0$ \cite{prb-polyac}.
The results presented in this paper were obtained by keeping $m=20$ states
in each renormalized block. This would correspond to
$4^4m^3 = 2,048,000$ possible configurations. However, only between 
$100,000$--$130,000$ of them belong to the targeted $S_z = S$ subspace. 
Still, the dimension of the superblock density-matrix, $16m^2=6400$,
is quite important, which renders the computations very demanding.

\section{Results and discussion}
\label{sec:res}

\begin{figure}[x]
\vspace{1cm}
\centerline{\resizebox{7cm}{7cm}{\includegraphics{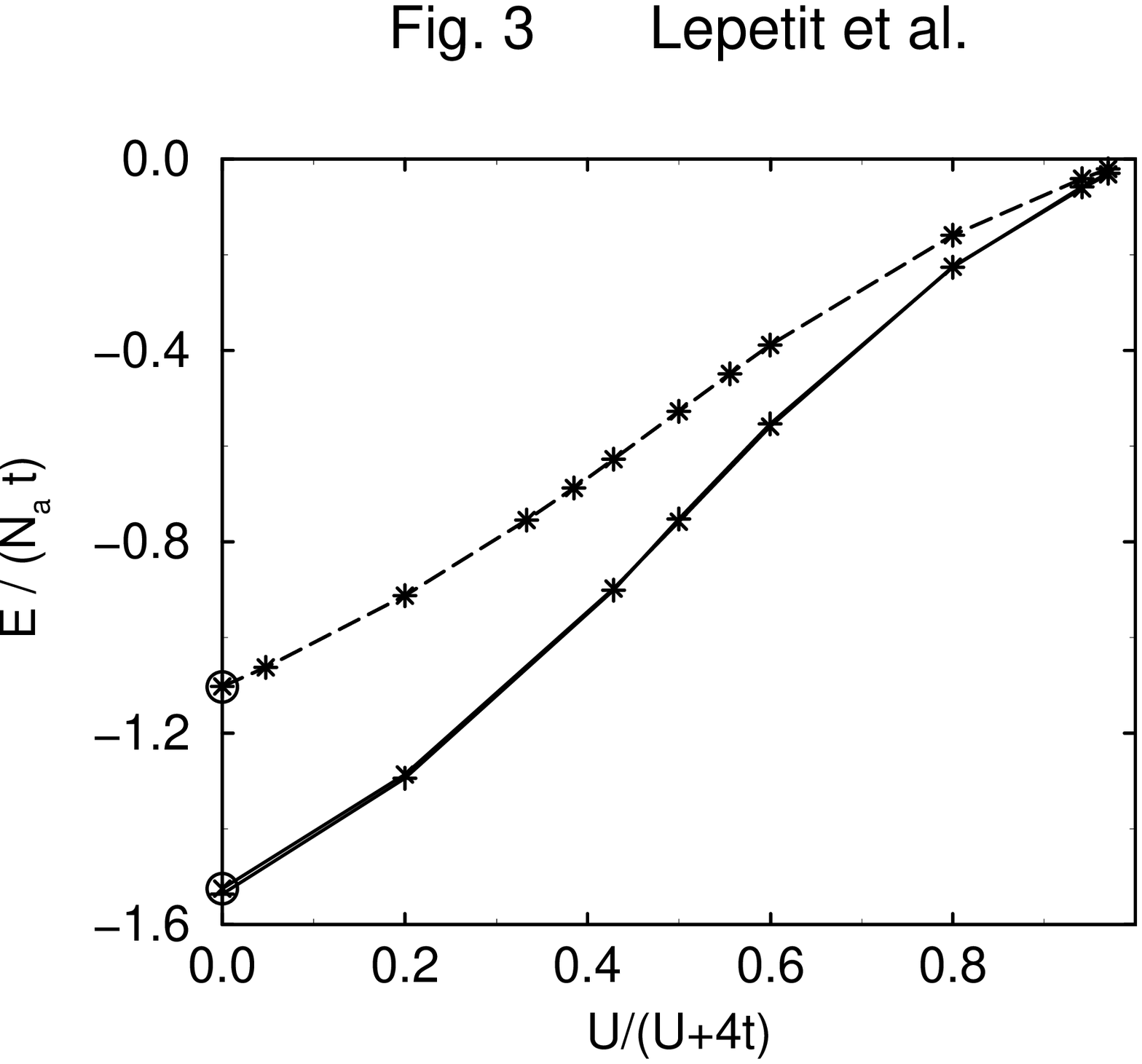}}}
\vspace{5mm}
\caption{
Ground-state energy per site of the half-filled Hubbard model on
a $z=3$ Bethe lattice as a function of the Coulomb repulsion $U/t$.
The solid curve is calculated locally at the central sites from 
$E= U \langle  \hat n_{0\uparrow} \hat n_{0\downarrow}\rangle + 
(zt/2) \sum_\sigma \langle \hat c_{1\sigma}^\dagger \hat c_{0\sigma} +
         \hat c_{0\sigma}^\dagger \hat c_{1\sigma}\rangle$ 
(see Fig.~\protect\ref{fig:bethe-lat1}). 
The dashed curve is the average energy $E_s$ of the complete system 
including the surface. Crosses (plus signs) correspond to even (odd) 
renormalization iterations. The corresponding exact results for $U=0$ are 
given by the open circles.}
\label{fig:ener}
\end{figure}

In Fig.~\ref{fig:ener} results are given for the ground-state energy 
per site as a function of $U/t$. 
Since in the Bethe lattice the weight 
of the atoms close to the surface does not vanish with increasing 
number of shells, it necessary to discern between global properties
which include surface contributions and local properties calculated 
close to the central site $i=0$. The global energy $E_s$ (dashed curve) 
is derived by extrapolating the ground-state energy per site
for $L\to\infty$. For $U=0$ the DMRG calculations
yield $E_s=-1.10268t$, while the exact results obtained by diagonalizing 
the finite-$L$ tight-binding matrix and extrapolating to $L\to\infty$
is $E_s^{ex} = -1.10306t$ (see Appendix). 
The agreement seems quite remarkable since $E_s$ is given by the contributions
of the atoms of the outermost shells ---~1/2 of the sites belong to the
surface, 1/4 to the layer below the surface, etc.~--- which are renormalized
already from the very first iterations. Similarly good results
are obtained for the local ground-state energy $E$ (solid curve)
which is obtained from the density matrices at the central sites
extrapolated for $L\to\infty$. 
For $U=0$, we obtain $E = -1.5247t$, while the integral of the single-particle
density of states of the Bethe lattice is $E^{ex}= -1.5255t$. The 
renormalized blocks thus provide a proper embedding of the central sites.
It is worth noting that the accuracy of these results, derived by 
keeping only $m=20$ states per block and setting $S_z=S$ at each 
iteration, is comparable to the accuracy of calculations with 
$m=100$--$150$ in the 1D Hubbard model ($S_z=0$).
It goes without saying that $m=100$ calculations on a Bethe lattice 
are hardly feasible with present computer facilities.
Since the uncorrelated limit is the most difficult (less precise) 
case in DMRG calculations on the Hubbard model \cite{prb-polyac}, 
we may expect that the results for finite $U/t$ are at least as accurate. 
This is also confirmed by the sum $P_m$ of the retained eigenvalues of the 
block density-matrix which increases with increasing $U/t$.

$E$ and $E_s$ increase monotonously with $U/t$ and vanish as expected 
in the Heisenberg limit. Their $U/t$ dependence are very 
similar. In fact, allowing for a rescaling of the energies at $U=0$, 
$E/E(U\!=\!0)$ and $E_s/E_s(U\!=\!0)$ are close to the corresponding 
results for the 1D Hubbard chain \cite{lieb-wu}. Quantitatively, $E_s$
is always higher than $E$ due to surface boundary effects. Surface sites 
at the outermost shell have a smaller local coordination number $z=1$.
Therefore the effective local band width and the binding energy of surface
sites are smaller than in the bulk.

\begin{figure}[x]
\vspace{1cm}
\centerline{\resizebox{7cm}{7cm}{\includegraphics{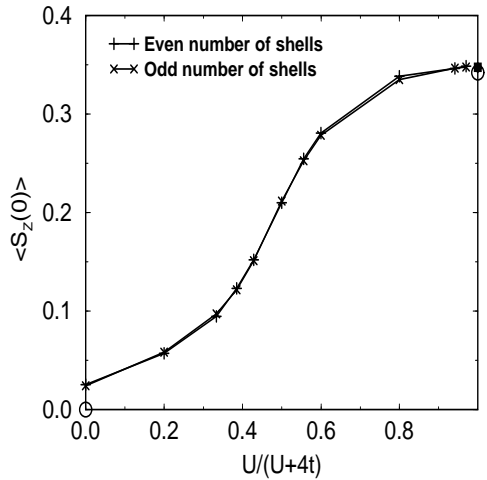}}}
\vspace{5mm}
\caption{
Average magnetization $\langle \hat S_z(0)\rangle$ for the half-filled Hubbard 
model on a Bethe lattice ($z=3$) calculated at the central site $i=0$.
Circles indicate tight-binding exact results ($U=0$) or
perturbation theory estimates ($U=\infty$). The dot is the 
DMRG result for the Heisenberg model on the $z=3$ Bethe lattice. 
        }
\label{fig:Sz}
\end{figure}

Several local properties have been calculated around the central site $i=0$
in order to analyze the behavior of the ground-state in the bulk limit. 
Results for the average local magnetic moment $\langle \hat S_z(0)\rangle$ 
are shown in Fig.~\ref{fig:Sz}.

$\langle \hat S_z(0)\rangle$ increases monotonously with $U/t$ reaching 
$0.35$ in the Heisenberg limit. This reflects the formation of a 
spin-density wave (SDW) in which the average spins on the two 
sublattices $A$ and $B$ point in opposite directions. Indeed, 
for NN sites ($i=0$ and $i=1$) we obtain 
$\langle \hat S_z(1)\rangle \simeq - \langle \hat S_z(0)\rangle$ and 
$\langle \hat S_z(1')\rangle \simeq  \langle \hat S_z(1)\rangle$ 
($|\langle \hat S_z(0) + \hat S_z(1)\rangle|\le 10^{-4}$).  
Notice that the calculated $\langle \hat S_z(0)\rangle$ 
does not vanish in the uncorrelated limit as it should. From the tight-binding
solution for the Bethe lattice with a finite number of shells $L$ 
one obtains $\langle \hat S_z(0)\rangle = 1/(3L+2)$ for $L$ even and 
$\langle \hat S_z(0)\rangle = 0$ for $L$ odd (see the Appendix). 
Thus, in finite-size systems with even $L$ the local magnetization 
does not vanish even for $U=0$ ($S_z=S$). 
Our DMRG calculations follow precisely the exact results during the 
first renormalization iterations but for large $\nu$ a slight increase of
$\langle \hat S_z(0)\rangle$ is observed for $L$ odd which yields 
a small non-zero value in the extrapolation to $L\to \infty$
($\langle \hat S_z(U\!=\!0)\rangle = 0.025$).
We expect that this inaccuracy will be remedied by increasing 
the number of states kept in each renormalized block. In any case, since 
the DMRG method has better convergence properties with increasing 
$U/t$, the precision of the results improves rapidly for finite 
values of the interaction. In fact, in the strongly correlated limit
the calculated $\langle S_z(0)\rangle = 0.35$ is in very good agreement
with perturbative estimations and with DMRG calculations using the Heisenberg 
Hamiltonian.

In order to analyze the strongly interacting Heisenberg
limit we approximate the ground-state wave function by including 
first-order perturbations to the N\'eel state $\phi_0$:
\begin{equation}
\psi^{(1)} = \phi_0 - {1 \over 2(z-1)}
\sum_{\langle i,j\rangle} (S_j^+ S_i^- + S_j^- S_i^+ ) \phi_0\; .
\label{eq:psi1}
\end{equation}
Notice that the coefficient of the first-order correction (spin-flip states) 
is independent of the exchange constant $J=4t^2/U$, since the 
off-diagonal matrix elements ($J/2$) and the energy differences 
[$2(z-1)J/2$] are both proportional to $J$ in the Heisenberg model.
The average of local operators $\hat O$ [e.g., $\hat O =\hat S_z(0)$ or 
$\hat O =\hat S_z(0)\hat S_z(1)$] are obtained from
\begin{equation}
\langle \hat O \rangle = { {\rm Tr} [\hat \rho   \hat O] 
\over {\rm Tr} [\hat \rho ] } \; ,
\label{eq:avo}
\end{equation}
where $\hat \rho$ refers to the reduced density matrix. For example, 
for a single site Eq.~(\ref{eq:psi1}) yields 
$\rho(\uparrow, \uparrow) = 1$, $\rho(\downarrow,
\downarrow) = 3/16$ and $\rho(\uparrow, \downarrow) =
\rho(\downarrow,\uparrow)=0$. In this way one obtains $\langle \hat
S_z(i) \rangle = 13/38=0.342$, which compares very well with the DMRG
result $\langle \hat S_z(i)\rangle = 0.348$ for $U/t =128$.  We
conclude that the reduction of $\langle \hat S_z(i)\rangle$ with
respect to the saturation value 1/2 at very large $U/t$ is the result
of quantum spin fluctuations involving the first NN shell of atom $i$.
In addition, we have also performed DMRG calculations for the
Heisenberg model on the $z=3$ Bethe lattice which confirm the Hubbard
results at $U/t\gg 1$  [$\langle \hat S_z(i)\rangle=0.347$, see
Fig.~\ref{fig:Sz}].

\begin{figure}[x]
\vspace{1cm}
\centerline{\resizebox{6cm}{6cm}{\includegraphics{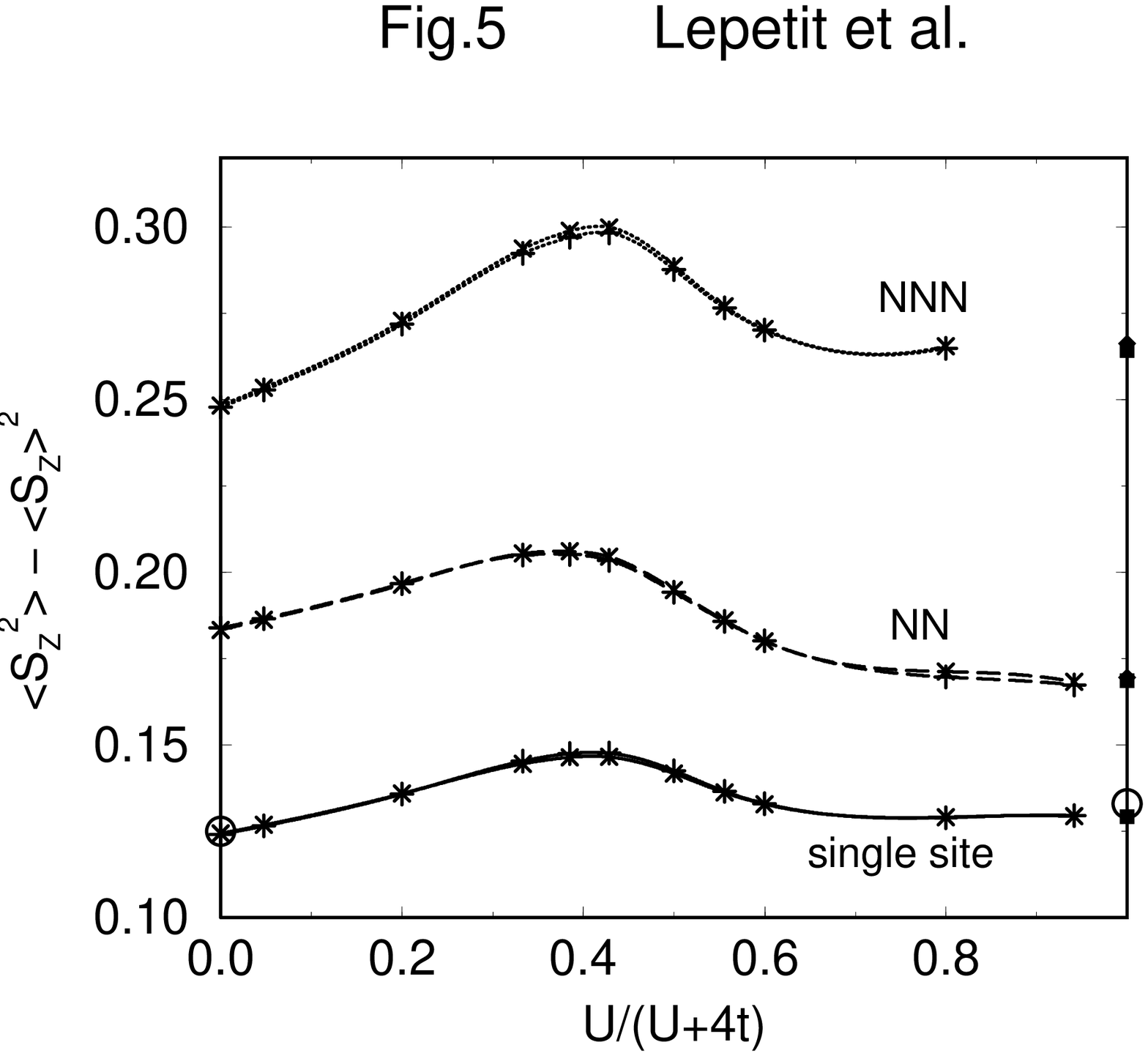}}}
\vspace{5mm}
\caption{
Bethe-lattice results for the average spin fluctuations 
$\langle\hat  S_z^2\rangle - \langle\hat  S_z\rangle^2$ as a function $U/t$,
where $\hat S_z = \hat S_z(0)$ (single site, solid curve),
$\hat S_z = \hat S_z(0) + \hat S_z(1)$ (NN spins, dashed curve) and
$\hat S_z = \hat S_z(1) + \hat S_z(1')$ (NNN spins, dotted curve).
See Fig.~\protect\ref{fig:bethe-lat1}. 
Circles indicate tight-binding exact results ($U=0$) or
perturbation theory estimates ($U=\infty$). The dots are
DMRG results for the Heisenberg model on the $z=3$ Bethe lattice. 
        }
\label{fig:vsz}
\end{figure}
\begin{figure}[x]
\centerline{\resizebox{6cm}{6cm}{\includegraphics{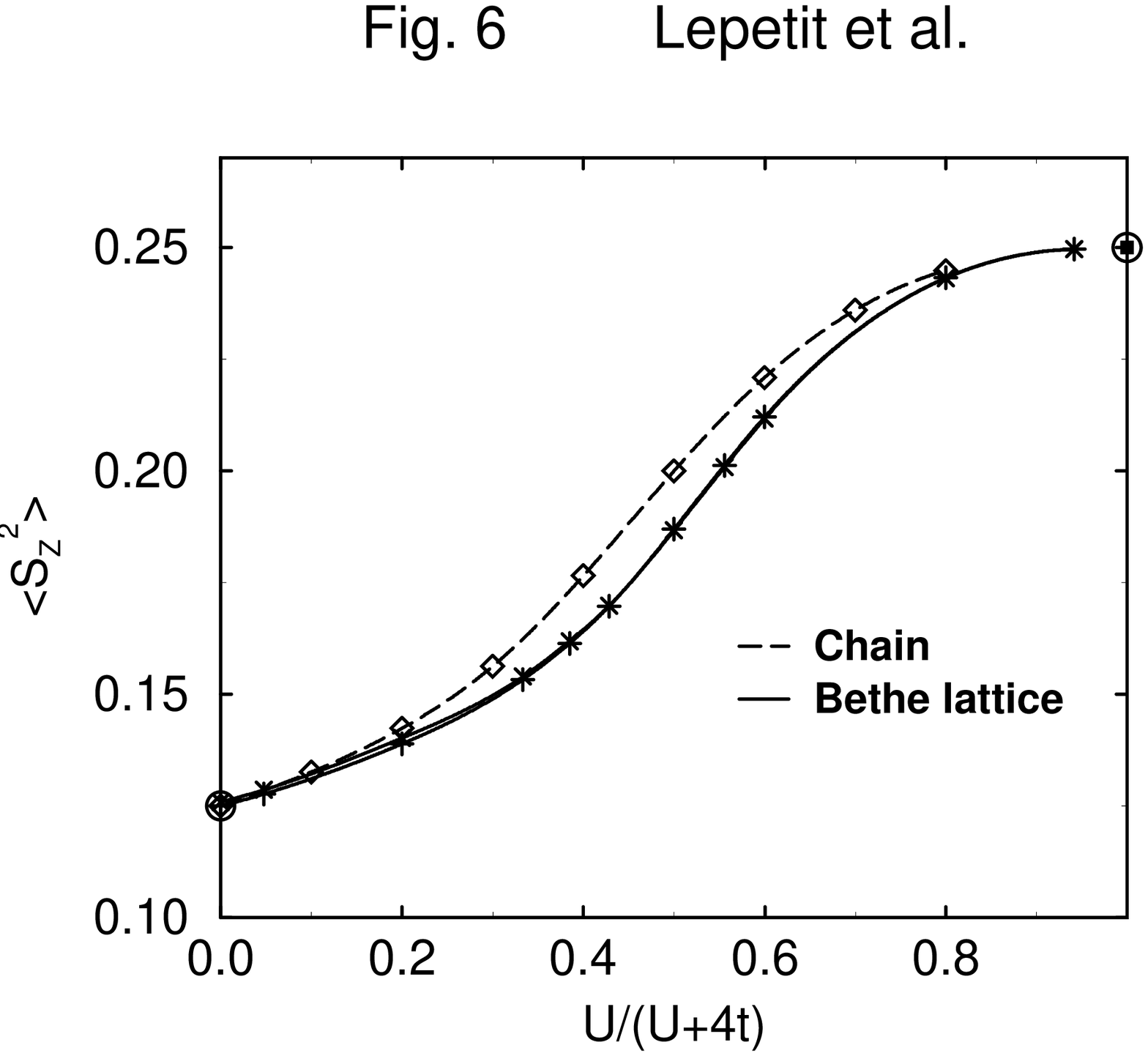}}}
\vspace{5mm}
\caption{
Results for $\langle \hat S_z(0)^2 \rangle$ of the half-filled Hubbard model
as a function $U/t$. The solid curve corresponds to the $z=3$ Bethe lattice
and the dashed curve to the one-dimensional chain ($z=2$).
        }
\label{fig:Sz2}
\end{figure}
In Fig.~\ref{fig:vsz} results are given for the fluctuation of the
local magnetic moments at a single site and at pairs of NN and NNN
sites.  In all cases $\langle S_z^2\rangle - \langle S_z\rangle^2$
presents a maximum for $U/t=2.4$--$2.9$. This behavior is a
consequence of the interplay between the $\langle S_z^2\rangle$
increase which is due to the reduction of the weight of doubly
occupied and empty sites by correlations, and the formation of
permanent magnetic moments $\langle S_z\rangle$, which reduces the
fluctuation of the spin moments around their average (see
Fig.~\ref{fig:Sz}).  As shown in Fig.~\ref{fig:Sz2}, $\langle
S_z^2\rangle$ in the Bethe lattice increases monotonously with $U/t$
very much like in the 1D Hubbard chain. The main qualitative
difference between Bethe-lattice and 1D results for $\langle
S_z^2\rangle - \langle S_z\rangle^2$ comes from $\langle S_z\rangle$
which is zero for the 1D case.  Notice that the DMRG calculations are
in good quantitative agreement with the tight-binding analytic results
(open circles, $U=0$), with independent DMRG calculations for the
spin-$1/2$ Heisenberg model (dots, $U=\infty$) as well as with the
first-order-perturbation estimation from Eqs.~(\ref{eq:psi1}) and
(\ref{eq:avo}) (open circles, $U=\infty$).

\begin{figure}[x] 
\vspace{1cm}
\centerline{\resizebox{7cm}{7cm}{\includegraphics{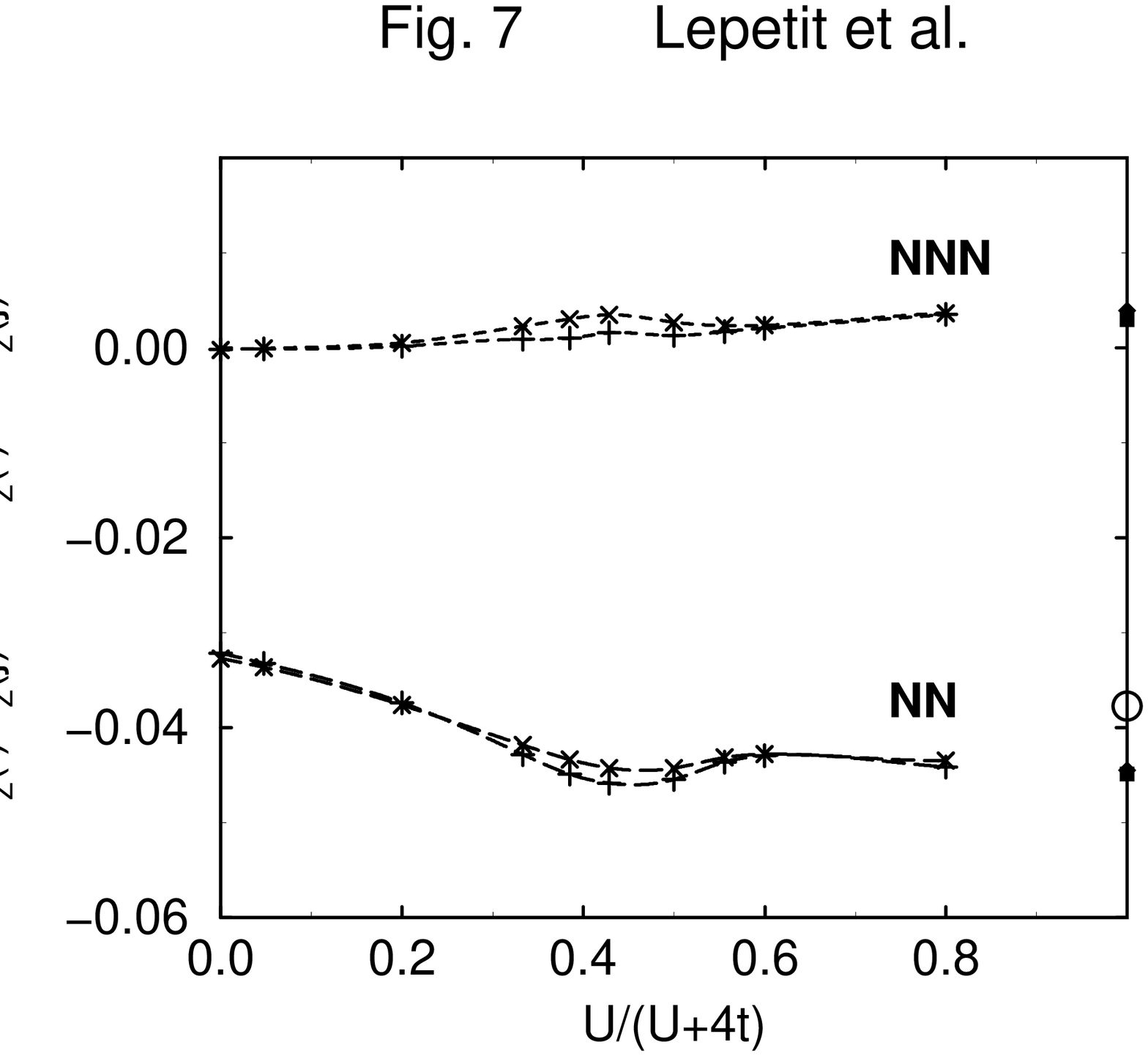}}}
\vspace{5mm}
\caption{
Spin correlation functions  $\langle \hat S_z(i) \hat S_z(j)\rangle -
\langle \hat S_z(i)\rangle \langle \hat S_z(j)\rangle$ between 
NN sites (dashed curve) and NNN sites (dotted curve).
As in Fig.~\protect\ref{fig:vsz}, 
circles indicate tight-binding exact results ($U=0$) or
perturbation theory estimates ($U=\infty$) and dots are 
DMRG results for the Heisenberg model.
}
\label{fig:flcorr} 
\end{figure}
The spin correlation functions $\langle \hat S_z(i) \hat S_z(j)\rangle -
\langle \hat S_z(i)\rangle \langle \hat S_z(j)\rangle$ between 
NN and NNN sites $i$ and $j$ are given in Fig.~\ref{fig:flcorr}.
NN spins show antiferromagnetic correlations which tend to become 
stronger as $U/t$ increases. A shallow maximum is observed
approximately at the same value of $U/t$ for which 
$\langle S_z^2\rangle - \langle S_z\rangle^2$ is maximal. 
Notice that a considerable part of the antiferromagnetic correlations 
between NN is already present for $U=0$. In contrast, the spin correlations
between NNN's are much weaker. In this case,
parallel alignment is slightly favored as $U/t$ increases 
[$\langle \hat S_z(1) \hat S_z(1')\rangle -
\langle \hat S_z(1)\rangle \langle \hat S_z(1')\rangle > 0$]. 
These trends are consistent with the sign alternations found in
$\langle \hat S_z(i)\rangle$ for $i$ belonging to different sublattices
(SDW). As expected, good agreement is obtained between the Hubbard results 
for large $U/t$ and independent DMRG calculations for the the Heisenberg 
model. The first-order estimates [Eq.~(\ref{eq:psi1})] are somewhat 
less accurate in this case but still remain qualitatively correct.

\begin{figure}[x]
\vspace{1cm}
\centerline{\resizebox{7cm}{7cm}{\includegraphics{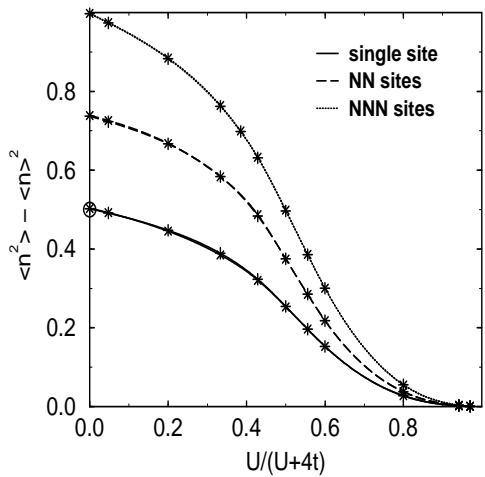}}}
\vspace{5mm}
\caption{
Bethe-lattice results for the density fluctuations 
$\langle \hat n^2\rangle - \langle \hat n\rangle^2$ as a function $U/t$,
where $\hat n = \hat n(0)$  (single site, solid curve),
$\hat n = \hat n(0) + \hat n(1)$ (NN sites, dashed curve) and
$S_z = \hat n(1) + \hat n(1')$ (NNN sites, dotted curve).
$\hat n(i) = \hat n_{i\uparrow} + \hat n_{i\downarrow}$, 
see Fig.~\protect\ref{fig:bethe-lat1}.
        }
\label{fig:vne}
\end{figure}

The average density $\langle \hat n(i)\rangle$ at the central 
sites $i=0$ and $i=1$ is very close to 1 independently of $U/t$, 
which confirms 
the expected absence of a charge-density wave for $U\ge 0$
($|\langle \hat n_{i\uparrow} + \hat n_{i\downarrow}\rangle - 1|<10^{-4}$).
The fluctuations of the density 
at a single site and at pairs of NN and NNN sites are given 
in Fig.~\ref{fig:vne}. 
In all cases we observe 
a monotonic crossover from the uncorrelated regime
($\langle \hat n^2\rangle - \langle \hat n\rangle^2$ maximal) to the 
strongly correlated, localized regime where charge fluctuations are
suppressed. The $U/t$ dependence is quite similar to what is obtained 
for 1D Hubbard model.
The density fluctuation at a pair of NNN sites (dotted curve) is
approximately twice the single-site result (solid curve) which indicates,
as in the case of the spin degrees of freedom, that 
density correlations between NNN's are very weak
[$\langle \hat n(1) \hat n(1') \rangle - 
\langle \hat n(1)\rangle \langle \hat  n(1')\rangle \simeq 0$]. 
In contrast, for NN sites charge fluctuations are significantly 
smaller. One observes that
$\langle [\hat n(0) + \hat n(1)]^2 \rangle - 
\langle \hat n(0) + \hat  n(1)\rangle^2 \simeq 
(3/2) [\langle \hat n(0)^2  \rangle - \langle \hat n(0)\rangle^2]$
or equivalently 
$\langle \hat n(0) \hat n(1) \rangle - 
\langle \hat n(0)\rangle \langle \hat  n(1)\rangle \simeq 
[\langle \hat n(0)^2  \rangle - \langle \hat n(0)\rangle^2]/2$.
Notice that these relations hold approximately for all 
values of $U/t$, even for $U=0$. The ratio between single-site 
charge fluctuations and fluctuations on a pair of NN's is not much 
affected by changes in the Coulomb repulsion strength and therefore
seems to result mainly from the geometrical proximity of NN sites.

\section{Conclusion}
\label{sec:conc}

Several ground-state properties of the half-filled Hubbard model 
have been determined on the Bethe lattice with coordination $z=3$
by using a density-matrix renormalization group (DMRG) algorithm for
open infinite systems. Though the lattice is not 
one dimensional (1D), the existence of a
unique path between any pair of sites allows to formulate a simple
renormalization procedure. In contrast to previous density-matrix 
renormalizations studies of Hubbard-like models 
on 1D chains or ladders where the number of sites $N_a$ increases 
linearly with the number of iterations $\nu$, in the present approach 
$N_a$ increases exponentially with $\nu$ ($N_a = 3\times 2^\nu - 2$). 
This is a consequence of the fact that 2 blocks are renormalized 
into a single one at each iteration. Despite the
very rapid increase of $N_a$, the DMRG method provides accurate results
over the whole range of $U/t$ already by keeping few states 
per block ($m=20$). This is achieved by working in the subspace 
of maximal spin projection $S_z = S$, where the ground-state spin
$S=  2^{\nu-1}$ is derived from a theorem by Lieb \cite{liebS}. 
For example, in the limit of $U=0$ the calculated ground-state 
energy per site differs by only $3\times 10^{-4}t$ from the exact 
tight-binding result. It is remarkable that this level
of precision concerns not only local properties calculated at the 
unrenormalized central sites, but also global properties
which are dominated by the renormalized sites of the outermost shells.
From a general point of view, the present study encourages
renormalizations of more than one block into a superblock in 
future DMRG algorithms.

The main results for the $z=3$ Bethe lattice may be summarized as follows.
The calculated local magnetic moments $\langle \hat S_z(i)\rangle$ 
increase monotonically with increasing Coulomb repulsion $U/t$ 
forming an antiferromagnetic spin-density-wave state which matches
the two sublattices of the bipartite Bethe lattice.
The maximum $\langle \hat S_z(i)\rangle$ found in the 
Heisenberg limit ($\langle \hat S_z(i)\rangle = 0.35$) is reduced with 
respect to the saturation value $\langle \hat S_z(i)\rangle = 1/2$
as a result of exchange flips between the spin at site $i$ and its NN's
that point in opposite directions. The fluctuations of the local spins
$\langle S_z(i)^2\rangle - \langle S_z(i)\rangle^2$ 
show a maximum as a function of $U/t$ for $U/t=2.4$--$2.9$.
For small $U/t$ the usual increase of $\langle S_z(i)^2\rangle$ 
due to the reduction of double occupations dominates, while for large $U/t$
the formation of large permanent moments $\langle S_z(i)\rangle$
blocks local spin fluctuations. NN sites show important antiferromagnetic 
spin correlations which increase with increasing Coulomb repulsion.
However, NNN sites are very weakly correlated over the whole range
of $U/t$. The AF correlations are very short ranged
in contrast to the static picture of a SDW. This reflects the 
importance of quantum fluctuations in the $z=3$ Bethe lattice as in 
1D systems.

Taking into account that the Bethe lattice is one of the standard models 
for the studying the limit of infinite dimensions, it would be very 
interesting to compare the present results for $z=3$ with the outcome of 
the $d=\infty$ equations in order to quantify the importance of $1/d$ 
corrections. Moreover, further DMRG investigations including NNN hoppings 
should be very valuable for a better understanding of the 
metal-insulator transition in the presence of frustrations, particularly
since the information derived from the DMRG method is complementary 
to finite-temperature quantum Monte-Carlo calculations.

\acknowledgements
Computer resources provided by IDRIS (CNRS) under project No.~960806 
are gratefully acknowledged.

\newpage
\section*{Appendix}

\begin{figure}[x]
\vspace{1cm}
\centerline{\resizebox{4cm}{4cm}{\includegraphics{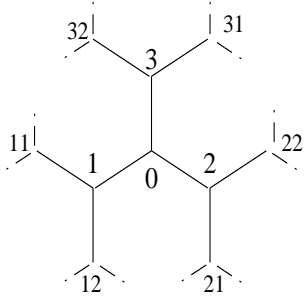}}}
\vspace{5mm}
\caption{Labelling of Bethe-lattice sites used in this appendix.}
\label{fig:bethe-appendix}
\end{figure}

The aim of this section is to outline the block diagonalization of 
the tight-binding matrix of a {\em finite} Bethe lattice formed by 
a central site $i=0$ and $L$ successive nearest neighbor (NN) shells.
The number of sites at shell $l$ ($1\le l\le L$) is given by
$N_s(l) = z (z-1)^{l-1}$, where $z$ refers to the coordination number. 
The notation used for labelling the lattice
sites is illustrated in Fig.~\ref{fig:bethe-appendix} for $z=3$.
A site belonging to the shell $l$ is denoted by the set of $l$ numbers 
$(i_1, i_2, \dots i_l)$ which define the path to follow in order to reach 
the desired site starting from the central site $i=0$.
The tight-binding matrix $H_0$ of the $z=3$ Bethe lattice with
NN hoppings $t$ can be block diagonalized by using that the finite 
Bethe lattice is invariant after the transposition of any pair of
branches that connect a site of shell $l-1$ with its 2 NN's of shell $l$.
For the outermost shell ($l = L$) the symmetry adapted single-particle
states are  
\begin{eqnarray}
|i_1, i_2, \dots i_{L-1}, + \rangle &=& 
\left( |i_1, i_2, \dots i_{L-1}, 1 \rangle +  |i_1, i_2, \dots i_{L-1}, 2
\rangle\right)/\sqrt{2} \\
|i_1, i_2, \dots i_{L-1}, - \rangle &=& 
\left( |i_1, i_2, \dots i_{L-1}, 1 \rangle -  |i_1, i_2, \dots i_{L-1}, 2
\rangle\right)/\sqrt{2} \; .
\end{eqnarray}
For the other shells ($l<L$) on proceeds recursively in the same way
building symmetric and antisymmetric linear combinations. For the first
shell the 3-fold symmetry around the central site $i=0$ is applied.

In the new basis $H_0$ splits into a $(L+1)\times(L+1)$ matrix block 
of the form 
\begin{equation}
A = \left[
\begin{array}{cccccc}
0          &\sqrt{3} t &  0        &\multicolumn{3}{c}{\cdots} \\
\sqrt{3} t &  0        &\sqrt{2} t & 0 & \multicolumn{2}{c}{\hfill} \\
0          &\sqrt{2} t &   0       &\sqrt{2} t&  \\
\vdots     &           & \ddots    & \ddots   & \ddots  \\
\multicolumn{5}{c}{\hfill} & \sqrt{2} t \\
\multicolumn{4}{c}{\hfill} & \sqrt{2} t & 0 
\end{array} \right] \; ,
\end{equation}
which involves only purely even states including the central site,
and in smaller $l\times l$ matrices $B_l$ with $1\le l\le L$ of the form
\begin{equation}
B_l = \left[
\begin{array}{cccccc}
0          &\sqrt{2} t &  0        &\multicolumn{3}{c}{\cdots} \\
\sqrt{2} t &  0        &\sqrt{2} t & 0 & \multicolumn{2}{c}{\hfill} \\
0          &\sqrt{2} t &   0       &\sqrt{2} t&  \\
\vdots     &           & \ddots    & \ddots   & \ddots  \\
\multicolumn{5}{c}{\hfill} & \sqrt{2} t \\
\multicolumn{4}{c}{\hfill} & \sqrt{2} t & 0 
\end{array} \right] \; .
\end{equation}
$B_L$ appears twice in $H_0$ and each of the other $B_l$ appears
$3\times 2^{L-l-1}$ instances in the total tight-binding matrix.

The eigenvalues of $B_l$ are $\beta_k = -2 \sqrt{2} t \cos{k \pi
\over l+1}$ with $k \in [1,l]$ and those of $A$ are
$\alpha_k = -2\sqrt{2}|t|\cos{\theta_k}$ where the $\theta_k$ are the roots
of $2 \sin{(L+2)\theta} = \sin{L\theta}$. 
The later equation is solved numerically and the tight-binding
energy per site $E_s$ of the finite Bethe lattice is determined for any $L$. 
The extrapolated value for $L\to \infty$ is $E_s = -1.10306t$. Notice that 
this result differs from the integral of the local density of states at the 
central site ($E = -1.5255t$) since $E_s(L\to \infty)$ is dominated by 
surface contributions.

%


\newpage


\begin{thebibliography}{99}


\bibitem{voll}
W. Metzner and D. Vollhardt, 
Phys. Rev. lett. {\bf 62}, 324 (1989);
D. Vollhardt, 
in {\em Correlated Electron Systems,} 
edited by V.J. Emery (World Scientific, Singapore, 1993).

\bibitem{kotliar}
M.J. Rozenberg, X.Y. Zhang and G. Kotliar, 
Phys. Rev. Lett. {\bf 69}, 1236 (1992);
M.J. Rozenberg, G. Kotliar and X.Y. Zhang, 
Phys. Rev. {\bf B49}, 10181 (1994).

\bibitem{georges}
A. Georges and W. Krauth, 
Phys. Rev. Lett. {\bf 69},1240 (1992);
Phys. Rev. B {\bf 48}, 7167 (1993).

\bibitem{hong}
J. Hong and H.Y. Kee, 
Phys. Rev. B {\bf 52}, 2415 (1995).

\bibitem{dmrg0}
S.R. White, Phys. Rev. Lett. {\bf 69}, 2863 (1992);
Phys. Rev. B {\bf 48}, 10345 (1993).

\bibitem{dmrg1}
S.R.\ White and D.A.\ Huse, Phys.\ Rev.\ {\bf B48}, 3844 (1993);
K.A.\ Hallberg, P.\ Horsh, G.\ Martinez, Phys.\ Rev.\ {\bf B52}, R719 (1995);
S.R.\ White, Phys.\ Rev.\ {\bf B53}, 52 (1996);
S.\ Liang and H.\ Pang, 
Europhys.\ Lett.\ {\bf 32}, 173 (1995);
S. Qin, S. Liang, Z. Su and L. Yu, 
Phys.\ Rev.\ B {\bf 52}, R5475 (1995);
B. Srinivasan, S. Ramasessha and H.R. Krishnamurthy, 
Phys.\ Rev.\ B {\bf 54}, 2276 (1996);
H. Pang and S. Liang, Phys. Rev. B {\bf 51}, 10287 (1995);
M.-B. Lepetit and G.M. Pastor, 
Phys. Rev. B {\bf 56}, 4447 (1997);
S.R.\ White, Phys.\ Rev.\ Lett.\ {\bf 69}, 2863 (1992);
C.C.\ Yu, S.R.\ White, Phys.\ Rev.\ Lett.\ {\bf 71}, 3866 (1993);
M.\ Guerrero, C.C.\ Yu, Phys.\ Rev.\ {\bf B51}, 10301 (1995).

\bibitem{otsuka}
H. Otsuka, 
Phys. Rev. B {\bf 53}, 14004 (1996).

\bibitem{hub}
J. Hubbard, 
Proc. R. Soc. London {\bf A276}, 238 (1963); {\bf A281}, 401 (1964); 
J. Kanamori, 
Prog. Theo. Phys. {\bf 30}, 275 (1963); 
M.C. Gutzwiller, 
Phys. Rev. Lett. {\bf 10}, 159 (1963).

\bibitem{liebS}
E.H. Lieb, 
Phys. Rev. Lett. {\bf 62}, 1201 (1989).

\bibitem{lieb-wu}
E.H. Lieb and F.Y. Wu, 
Phys. Rev. Lett. {\bf 20}, 1445 (1968).

\bibitem{prb-polyac}
M.-B. Lepetit and G.M. Pastor, 
Phys. Rev. B {\bf 56}, 4447 (1997).



\end{thebibliography}
\end{document}